\documentclass[12pt]{article}
\usepackage{graphicx}
\usepackage{bm}
\usepackage{amsmath}
\usepackage{amssymb}

\newcommand{\AM}{Annals Math. }
\newcommand{\CMP}{Commun. Math. Phys. }

\begin{document}
\title{Comment on 'Fisher information for quasi-one-dimensional hydrogen atom'}
\author{O. Olendski \thanks{Department of Applied Physics and Astronomy, University of Sharjah, P.O. Box 27272, Sharjah, United Arab Emirates}
}
\date{}
\maketitle

\begin{abstract}
A further argument is provided in the discussion on the correct form of the bound-state momentum wave function of the quasi-one-dimensional hydrogen atom; namely, considering its behavior at the large quantum indices, it is reconfirmed that the complex expression from Olendski (2017) {\it Eur. J. Phys.} {\bf 38} 038001 is a correct one. Groundlessness of other interpretations is also highlighted.
\end{abstract}
\vskip.7in
\maketitle
\noindent
A correct form of the bound-state momentum wave function $\Phi_n(p)$, $n=1,2,\ldots$, of the quasi-one-dimensional (Q1D) hydrogen atom (i.e., a quantum structure whose potential $V$ in position representation is given by
\begin{equation}\label{Q1Dpotential1}
V(x)=\left\{\begin{array}{cc}
-\frac{\alpha}{x},&x>0\\
\infty,&x\leq0,
\end{array}\right.
\end{equation}
$\alpha>0$) was derived in Ref.~\cite{Olendski1}. For doing this, its position counterpart $\Psi_n(x)$ was found first as a solution of the corresponding Schr\"{o}dinger equation, which yields the energy spectrum coinciding with the 3D hydrogen atom
\begin{equation}\label{Energies1}
E_n=-\frac{1}{2n^2},
\end{equation}
with the corresponding eigen functions represented as
\begin{equation}\label{SolutionX2}
\Psi_n(x)=\frac{2x}{n^{5/2}}e^{-x/n}L_{n-1}^{(1)}\!\left(\frac{2x}{n}\right),
\end{equation}
with $L_m^{(\beta)}(x)$, $m=0,1,\ldots$, being a generalized Laguerre polynomial \cite{Abramowitz1}. Starting from Eq.~\eqref{Energies1}, we use the Coulomb units where energies, distances and momenta are measured in terms of $m_p\alpha^2/\hbar^2$, $\hbar^2/(m_p\alpha)$, and $m_p\alpha/\hbar$, respectively, $m_p$ being a mass of the particle. Position and momentum waveforms are related through the Fourier transformation
\begin{equation}\label{Fourier1}
\Phi_n(p)=\frac{1}{\sqrt{2\pi}}\int_0^\infty e^{-ipx}\Psi_n(x)dx.
\end{equation}
As a result, one gets after simple but elegant calculation:
\begin{equation}\label{SolutionP1}
\Phi_n(p)=(-1)^{n+1}\sqrt{\frac{2n}{\pi}}\frac{(1-inp)^{n-1}}{(1+inp)^{n+1}}.
\end{equation}
Obviously, both $\Psi_n(x)$ and $\Phi_n(p)$ satisfy orthonormality requirements:
\begin{equation}\label{OrthoNormalityX1}
\int_0^\infty\Psi_{n'}(x)\Psi_n(x)dx=\int_{-\infty}^\infty\Phi_{n'}^\ast(p)\Phi_n(p)dp=\delta_{nn'},\\
\end{equation}
$\delta_{nn'}=\left\{\begin{array}{cc}
1,&n=n'\\
0,&n\neq n'
\end{array}\right.$ is a Kronecker delta, $n'=1,2,\ldots$. Observe that since the position function $\Psi_n(x)$ does not possess any symmetry, its momentum counterpart $\Phi_n(p)$ is essentially {\em complex}. One has to mention that the very similar form of the momentum waveform was derived for the motion on the whole $x$-axis in the potential $V(x)=-|x|^{-1}$ \cite{Nunez1}.

Comment \cite{Olendski1} appeared as a reaction to the resent research \cite{Saha1} where it was stated that the momentum function of the same structure is {\em real} and given in an awkward form involving Chebyshev polynomials  \cite{Abramowitz1} with the argument $(1-n^2p^2)/(1+n^2p^2)$. Authors of Ref.~\cite{Saha1} did not agree with the conclusions leading to Eq.~\eqref{SolutionP1}, and in their latest Arxiv submission \cite{Saha2} they found that their momentum waveform from Ref.~\cite{Saha1} is just an imaginary part of the function from Eq.~\eqref{SolutionP1}:
\begin{equation}\label{FunctionSTC1}
\Phi_n^{STC}(p)=\Im(\Phi_n(p))=(-1)^n\sqrt{\frac{2n}{\pi}}\frac{\sin(2n\arctan(np))}{1+n^2p^2}.
\end{equation}
Inspired by this discovery, Saha, Talukdar and Chatterjee (STC) \cite{Saha2} analyze the graphs of $\Re(\Phi_n(p))$ and $\Im(\Phi_n(p))$, which they borrow from Ref.~\cite{Olendski1}, with their main conclusion (provided without any proof or relevant references) stating that since the position wave function $\Psi_n(x)$ has $(n-1)$ zeros on the real semi axis $0\leq x<\infty$, its Fourier transform $\Phi_n(p)$ should exhibit similar behavior as a function of $p$. Moving forward, they declare a function from Eq.~\eqref{SolutionP1} 'physically inadmissible' since only its imaginary part vanishes at the origin while the real component, which, as was shown in Ref.~\cite{Olendski1}, is an even function of the momentum, has a nonzero value at $p=0$. Thus, STC state that only the function from Eq.~\eqref{FunctionSTC1} "exhibits  the  physically  acceptable number of nodes  for all values of $n$ and represents the true momentum-space wave  function for the Q1D hydrogen atom" \cite{Saha2}.
\begin{figure}
\centering
\includegraphics[width=\columnwidth]{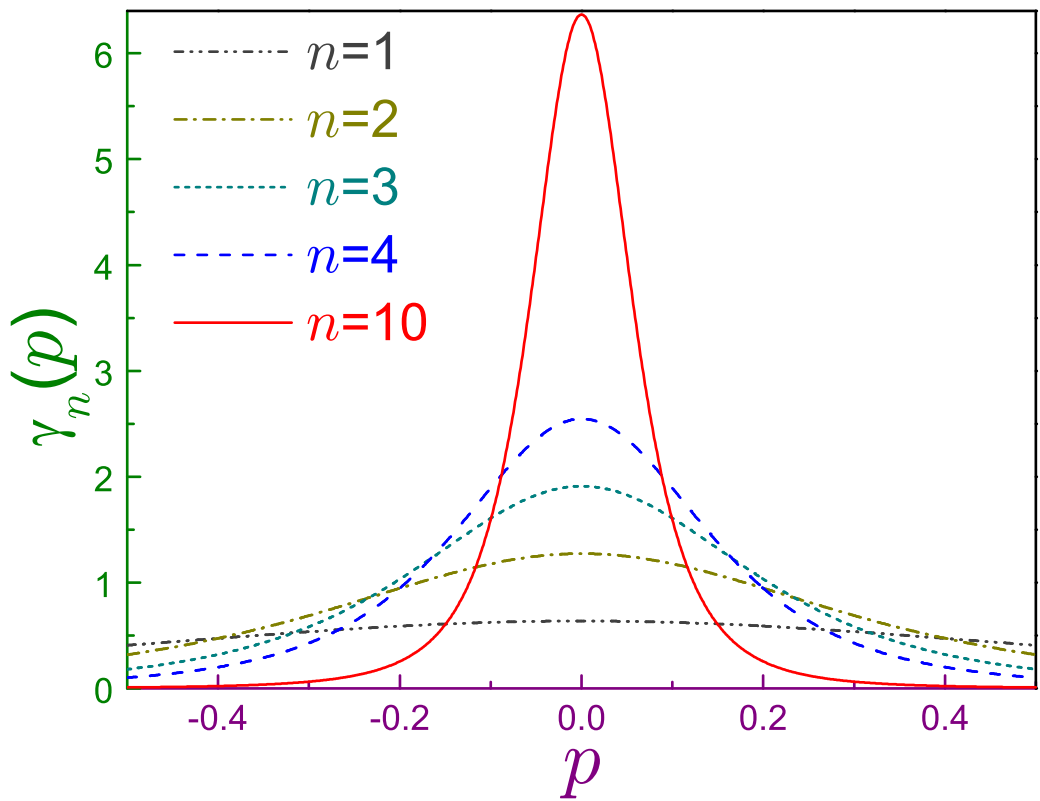}
\caption{\label{MomentumFig1}
Momentum densities $\gamma_n(p)$ where solid line is for $n=10$, dashed curve depicts the state with $n=4$, dotted line -- for $n=3$, dash-dotted curve -- for $n=2$, and dash-dot-dotted line -- for $n=1$.}
\end{figure}

Unfoundedness of such interpretation is evident and speaks for itself. To show a physical absurdity of these arguments, one has to recall that the most general physical meaning of the wave function is not determined by the number of its nodes and their correspondence to their counterparts in a conjugate space. The one and only fundamental physical meaning of the wave function is a square of its absolute value, which defines a probability of finding the quantum particle in the corresponding state. For the momentum density
\begin{equation}\label{Gamma1}
\gamma_n(p)=|\Phi_n(p)|^2
\end{equation}
one elementary finds for our system \cite{Olendski1}:
\begin{equation}\label{DensityP1}
\gamma_n(p)=\frac{2n}{\pi}\frac{1}{(1+n^2p^2)^2}.
\end{equation}
Consider what will happen in the limit of the large quantum numbers, $n\rightarrow\infty$. First of all, from Eq.~\eqref{Energies1} it follows that the corresponding energy turns to zero, $E_{n=\infty}=0$. Next, this quasi classical approximation implies that the energy is so high that the particle does not 'see' the corresponding potential and its total energy equals to the kinetic component only, $E_n=p^2$, $n\rightarrow\infty$, which, obviously, becomes zero too: $p\rightarrow0$  at $n\rightarrow\infty$. But this is exactly what the probability density from Eq.~\eqref{DensityP1} tells us; namely, the corresponding function at the quantum index $n$ tending to infinity transforms, as it is elementary to check, into the $\delta$-function:
\begin{equation}\label{Limit1}
\gamma_n(p)\rightarrow\delta(p),\quad n\rightarrow\infty.
\end{equation}
Thus, in this limit all moments but $p=0$ are eliminated. Everything is physically logical and consistent! Evolution of the density $\gamma_n(p)$ is shown in Fig.~\ref{MomentumFig1} for several indices. It is seen that the bell-like shape of $\gamma_n(p)$ at the larger $n$ sharpens around $p=0$. In turn, the cornerstone of the STC theory was the fact that the momentum function (and, accordingly, the corresponding probability density) is an identical zero at the zero momentum what strictly forbids the states with $p=0$ at any quantum index. This contradiction of the unphysical and mathematically faulty STC arguments with the quasi-classical limit reconfirms the correctness of Eq.~\eqref{SolutionP1}.

To make the function from Eq.~\eqref{FunctionSTC1} normalizable, STC multiply it by two; as a result, their density becomes:
\begin{equation}\label{FunctionSTC2}
\gamma_n^{STC}(p)=\frac{8n}{\pi}\frac{\sin^2(2n\arctan(np))}{(1+n^2p^2)^2}.
\end{equation}
The point is that this density is normalizable for the non negative $p$ only:
\begin{equation}\label{FunctionSTC3}
\int_0^\infty\gamma_n^{STC}(p)dp=1.
\end{equation}
STC do not explain why they do arbitrarily forbid motion with the negative momenta but, doing so, they commit another blunder; namely, the function from Eq.~\eqref{FunctionSTC2} violates Shannon entropy inequality \cite{Bialynicki1,Beckner1}:
\begin{subequations}\label{Shannon1}
\begin{align}\label{Shannon1total}
S_\rho+S_\gamma\geq1+\ln\pi
\intertext{with}
\label{Shannon1X}
S_\rho=-\int\Psi^2(x)\ln(\Psi^2(x))dx\\
\label{Shannon1P}
S_\gamma=-\int\gamma(p)\ln(\gamma(p))dp
\end{align}
\end{subequations}
and integrations carried out over all possible intervals. Calculating position Shannon entropy for the lowest level yields:
\begin{equation}\label{Shannon2X}
S_{\rho_1}=2\gamma=1.1544
\end{equation}
with $\gamma$ being Euler's constant:
\begin{equation}\label{Euler1}
\gamma=\lim_{n\rightarrow\infty}\left(\sum_{i=1}^n\frac{1}{i}-\ln n\right)=0.5772.
\end{equation}
Momentum entropy with  the correct density from Eq.~\eqref{DensityP1} and with the correct limits $-\infty<p<\infty$ can be evaluated analytically for any quantum state:
\begin{equation}
S_{\gamma_n}=-\ln\frac{2n}{\pi}+4\!\left(\ln2-\frac{1}{2}\right),
\end{equation}
yielding, in particular, for the ground level:
\begin{equation}
S_{\gamma_1}=1.2242,\quad S_{\rho_1}+S_{\gamma_1}=2.3786,
\end{equation}
satisfying, of course, Eq.~\eqref{Shannon1total} since
\begin{equation}
1+\ln\pi=2.1447.
\end{equation}
At the same time, STC density, Eq.~\eqref{FunctionSTC2}, multiplied by its own logarithm and integrated over non negative momenta only gives:
\begin{equation}
S_{\gamma_1}^{STC}=0.5575,\quad S_{\rho_1}+S_{\gamma_1}^{STC}=1.7119,
\end{equation}
in obvious violation of Eq.~\eqref{Shannon1total}. This is not surprising since the rigorous derivation of this inequality \cite{Bialynicki1,Beckner1} uses full Fourier transform, Eq.~\eqref{Fourier1}, and not its arbitrary castrated sine counterpart.

I would like to finish by citing again an except from Ref.~\cite{Saha2}; this time, the phrase from its last sentence will be used; namely, let us hope that the analysis provided above will help "... to resolve any confusion regarding the choice of momentum-space wave function for the Q1D hydrogen atom".


\begin{thebibliography}{99}
\bibitem{Olendski1}O. Olendski, {\it Eur. J. Phys.} {\bf 38}, 038001 (2017).
\bibitem{Abramowitz1}M. Abramowitz and I. A. Stegun  1964 \textsl{\textit{Handbook of Mathematical Functions }} (New York: Dover)
\bibitem{Nunez1}H. N. N\'{u}\~{n}ez-Y\'{e}pez, A. L. Salas-Brito, and D. A. Solis, {\it Phys. Rev. A} {\bf 83}, 064101 (2011).
\bibitem{Saha1}A. Saha, B. Talukdar, and S. Chatterjee, {\it Eur. J. Phys.} {\bf 38}, 025103 (2017).
\bibitem{Saha2}A. Saha, B. Talukdar, and S. Chatterjee, arXiv:1703.03578.
\bibitem{Bialynicki1}I. Bia{\l}ynicki-Birula and J. Mycielski, {\it\CMP} {\bf 44}, 129 (1975).
\bibitem{Beckner1}W. Beckner, {\it\AM} {\bf 102}, 159 (1975).
\end{thebibliography}
\end{document}